\documentclass[a4paper,11pt]{article}
\usepackage{pos}
\usepackage{booktabs}
\usepackage{gensymb}

\title{Confronting observations of VHE gamma-ray
blazar flares with reconnection models}
 \ShortTitle{Comparing blazar flares with reconnection models}

\author*[a]{J. Jormanainen}
\author[a,b]{T. Hovatta}
\author[a]{E. Lindfors}
\author[c]{I. Christie}
\author[d]{M. Petropoulou}
\author[a]{I. Liodakis}

\affiliation[a]{Finnish Centre for Astronomy with ESO, University of Turku, Finland}

\affiliation[b]{Aalto University Mets\"ahovi Radio Observatory, Mets\"ahovintie 114, FI-02540 Kylm\"al\"a, Finland}

\affiliation[c]{Northrop Grumman Corporation, 2980 Fairview Park Drive, Falls Church, VA 22042, USA}

\affiliation[d]{Department of Physics, National \& Kapodistrian University of Athens, University Campus, Zografos, 15784, Greece}

\emailAdd{jesojo@utu.fi}

\abstract{Several models have been suggested to explain the fast gamma-ray variability observed in blazars,  but its origin is still debated. One scenario is magnetic reconnection, a process that can efficiently convert magnetic energy to energy of relativistic particles accelerated in the reconnection layer.  In our study, we compare results from state-of-the-art particle-in-cell simulations with observations of blazars at Very High Energy (VHE, E > 100 GeV) gamma-rays. Our goal is to test our model predictions on fast gamma-ray variability with data and to constrain the parameter space of the model, such as the magnetic field strength of the unreconnected plasma and the reconnection layer orientation in the blazar jet. For this first comparison, we used the remarkably well-sampled VHE gamma-ray light curve of Mrk 421 observed with the MAGIC and VERITAS telescopes in 2013. The simulated VHE light curves were generated using the observable parameters of Mrk 421, such as the jet power, bulk Lorentz factor, and the jet viewing angle, and sampled as real data. Our results pave the way for future model-to-data comparison with next-generation Cherenkov telescopes, which will help further constrain the different variability models.}

\FullConference{37$^{\rm{th}}$ International Cosmic Ray Conference (ICRC 2021)\\
		July 12th -- 23rd, 2021\\
		Online -- Berlin, Germany}

\begin{document}
\maketitle

\section{Introduction}
\label{intro}

Blazars are active galactic nuclei (AGN) whose jet is seen closely aligned with our line of sight \cite{Urry1995}. They are observed to be extremely variable in various time scales across the entire electromagnetic spectrum \cite{Marscher2008, Marscher2010, Jorstad2010, Ahnen2016, Nilsson2018}, but in the VHE gamma-rays the source of their variability is still largely unknown. Magnetic reconnection offers one possiblity to explain the fastest observed VHE variations consisting of flares of only some hours even down to some minutes \cite{Giannios2009,Giannios2013}. In our study, we consider the model presented in \cite{Christie2019}. According to it, non-thermal emission is produced in plasmoids, i.e. quasi-spherical structures forming in current sheets during magnetic reconnection. The emission produced in plasmoids can be strongly variable up to the highest gamma-ray energies. To simulate this process, they combined 2D particle-in-cell (PIC) simulations of relativistic reconnection with a time-dependent radiative transfer code, which we utilize also in our study.

\begin{figure}[ht]
\centering
\includegraphics[scale=0.5]{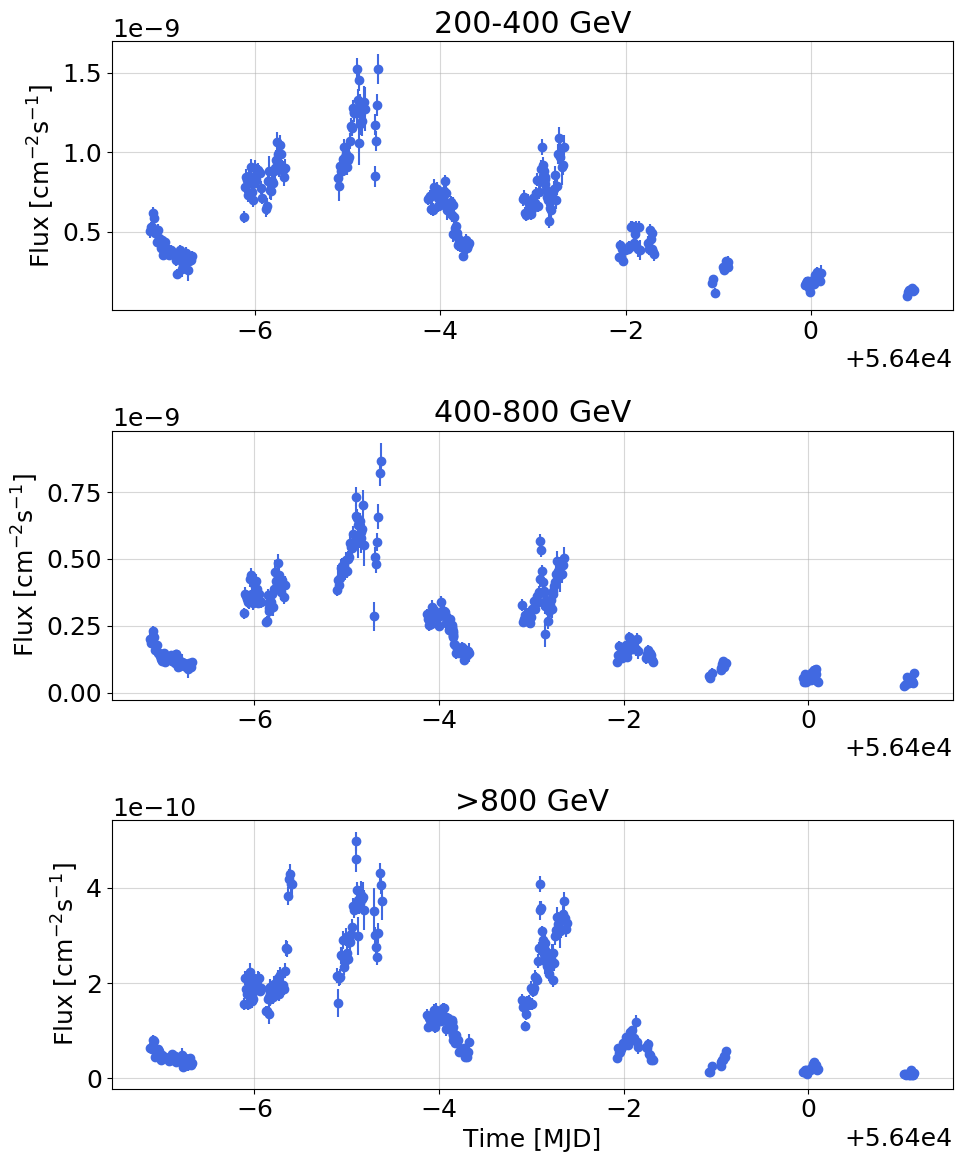}
  \caption{Real light curves of Mrk 421 in three energy bands of 200-400 GeV, 400-800 GeV, and >800 GeV observed by MAGIC and VERITAS in 2013 (\cite{Acciari2020}).}
  \label{fig:observed}
\end{figure}

In order to develop and introduce our method, we used only one particularly  well-sampled light curve of Mrk 421 (Fig.~\ref{fig:observed}) observed in a simultaneous campaign by MAGIC and VERITAS in VHE gamma-rays, and also by NuSTAR in X-rays in 2013 \cite{Acciari2020}. We use only the VHE data in our study. The data set consists of about 200 hours of observations in three energy bands, 200-400 GeV, 400-800 GeV, and >800 GeV, obtained during nine consecutive nights. The data observed by VERITAS have been scaled to match the level of MAGIC data as described in \cite{Acciari2020}. The analysis presented here can also be applied to light curves of other sources as well as other energies and time scales.

This proceedings article is structured as follows. In Section~\ref{sim}, we give a brief description of the simulation setup. In Section~\ref{analysis}, we explain the steps taken to compare the simulated and the observed data, and in Section~\ref{distr}, we highlight one of the developed analysis methods. In Section~\ref{results}, we state the findings of Section~\ref{distr} analysis, in Section~\ref{discussion}, we discuss the consequences of these results, and finally in Section~\ref{summary} we summarize our findings.

\section{Simulation setup}
\label{sim}

In order to compute the simulations individually for Mrk 421, observable parameters were collected from the literature if available. These parameters include the jet power $P_{jet}$, bulk Lorentz factor $\Gamma _{j}$, viewing angle of the jet $\theta _{obs}$, synchrotron peak frequency, $\gamma _{max}$ of the particle distribution, and magnetization of the jet $\sigma$. The values used for the simulation setup are given in Table~\ref{tab:table}. Ideally, these parameters help us to constrain the free parameters to realistic ranges and to obtain results that resemble the observed data. A more detailed description of the simulation setup is given in \cite{Christie2019}. Simulations of different jet scenarios were produced by altering the angle of the reconnection layer (current sheet) with respect to the jet axis between 0 and 180 degrees, the observing angle of the jet between 0 and 8 degrees, and for three different magnetic field strengths, B = 0.1 G, 1 G, 10 G, all in all, obtaining 285 different reconnection scenarios. The results presented in this paper are only for B = 0.1 G.

\begin{table}[ht]
\centering
\caption{Observed jet parameters collected from the literature for Mrk 421. }
\label{tab:table}
\begin{tabular}{@{}llllll@{}}
\toprule
log($P_{jet}$) [erg/s]  & $\Gamma _{j}$ & $\theta _{obs}$ [deg] & SED peak [Hz]  & $\gamma _{max}$   & $\sigma$ \\ \midrule
43.19 & 4     & 0-8     & 1.66E+016 & 9.5E+05 & 50    \\ \bottomrule
\end{tabular}
\end{table}

\section{Analysis}
\label{analysis}

\begin{figure}[ht]
\centering
\includegraphics[scale=0.5]{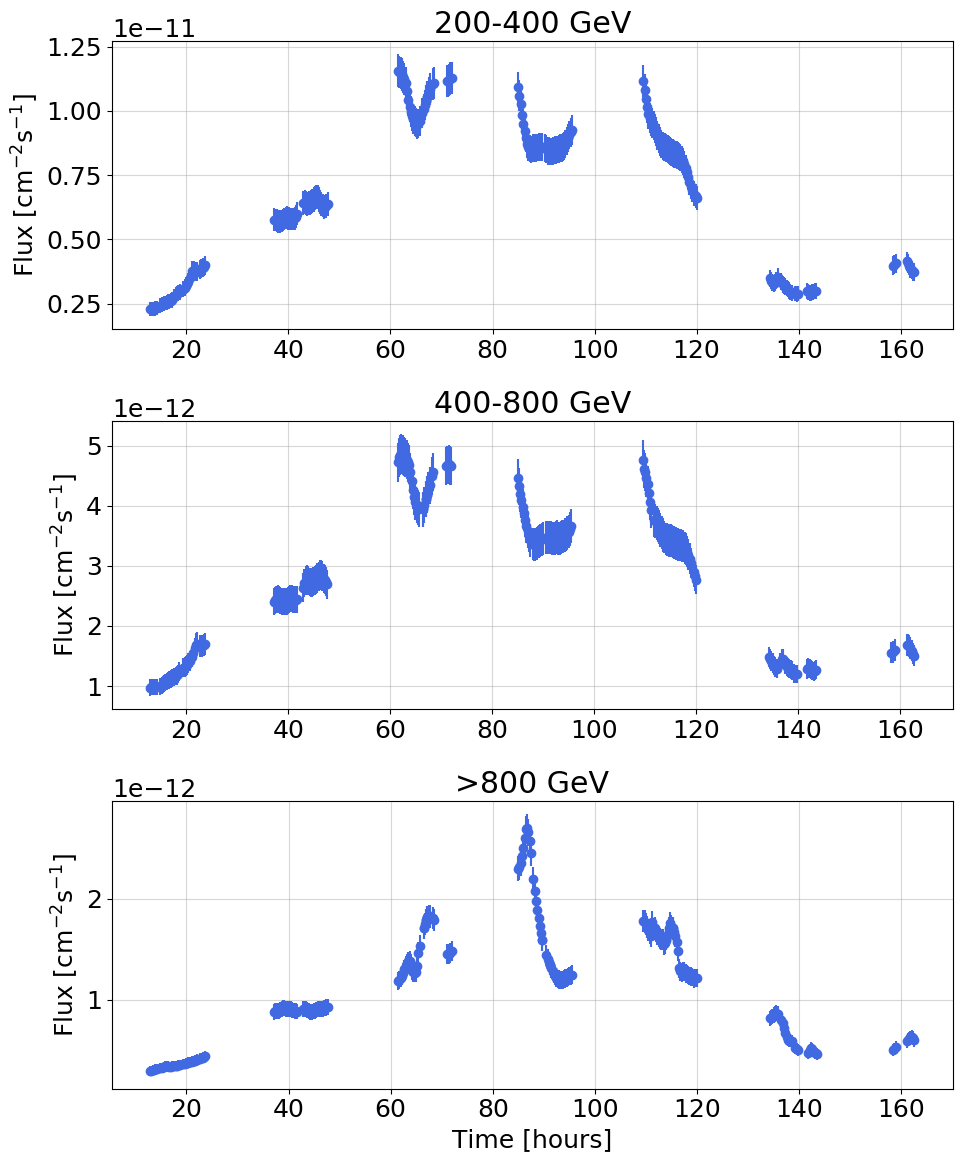}
  \caption{An example of a simulated light curve after the data has been treated for the comparison. This includes for example binning the data to 15-minute bins as well as introducing observational gaps.}
  \label{fig:simulated}
\end{figure}

Before the simulated light curves were compared with the observed data, they had to be treated in various ways to obtain light curves that resemble the observed ones. In observations, there are numerous things that prevent us from getting a continuous, precise signal from the source such as the visibility of the source throughout the day and year, readout times of the instruments, and different sources of error. All of these had to be taken into account before the comparison. The most important aspect of this treatment was producing fake light curves from simulated data by binning them with similar integration times as the observed data, and creating gaps in the data by utilizing the observed times of the real data. Figure~\ref{fig:simulated} shows an example of one simulated light curve after it has been treated to resemble the observed light curve.

In our analysis, we aim to combine several methods of comparing the simulated and the observed data sets in terms of the flux amplitudes and time scales. These methods include comparisons of flux distributions, fractional variability, and variability time scales of detected flares. The detailed description of each method is left for Jormanainen et al. (in prep.), but the flux distribution comparison is already introduced in the next Section.

\section{Results}
\label{results}

\subsection{Flux distribution comparison}
\label{distr}

\begin{figure}[ht]
\centering
\includegraphics[scale=0.5]{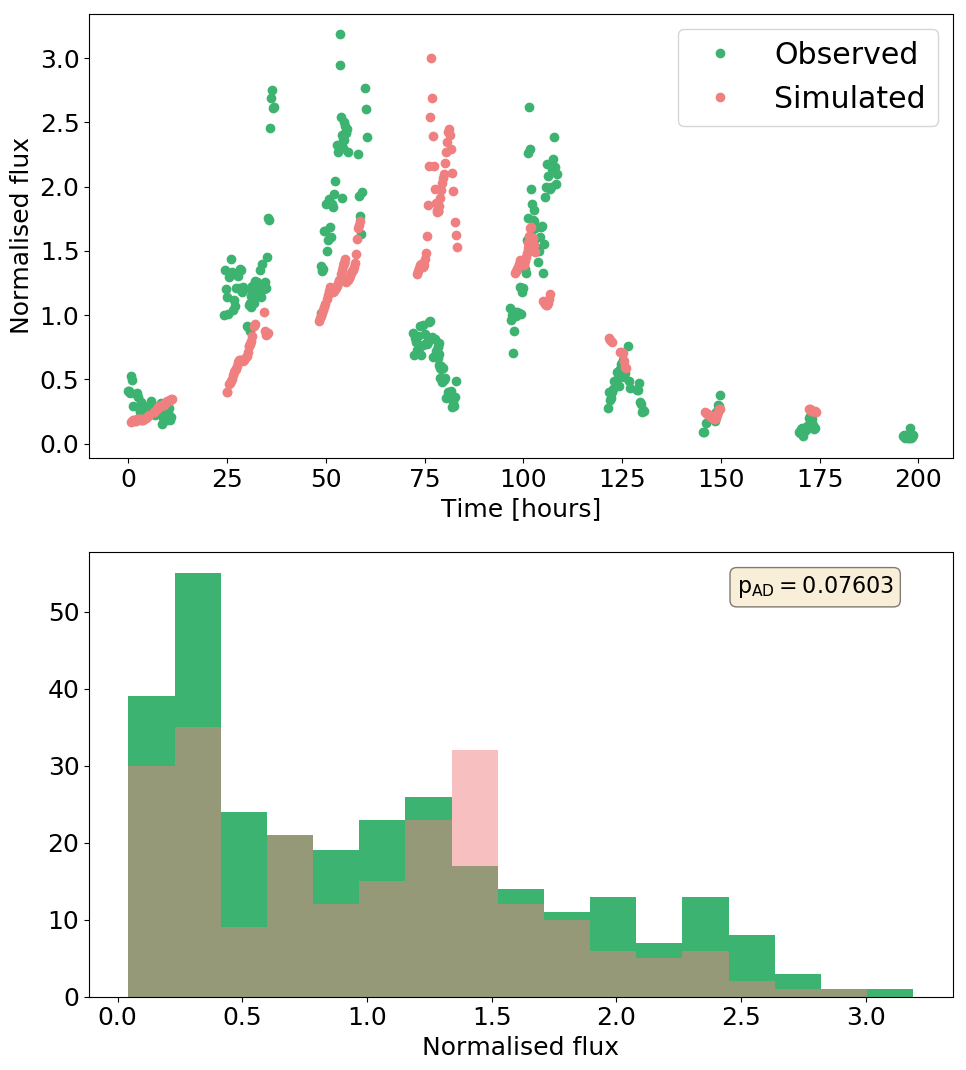}
  \caption{The upper panel shows a comparison of the normalized light curves, the observed data shown in green and the simulated data in pink. The lower panel shows the comparison of the flux distributions and the p-value of the Anderson-Darling test computed for the two distributions.}
  \label{fig:distr}
\end{figure}

Because our observed light curves include seasonal and daily gaps, we need to introduce these also into the simulated light curves. As the simulations were not always exactly the same length as the real data, and we do not know, which part of the simulated light curves would correspond to the \textit{observed portions}, the starting time of the observed times were shifted randomly to sample the full range of the simulated light curve. This was repeated a thousand times for one simulation scenario to obtain 1000 "observations" of one simulated light curve. Because the simulated fluxes were often 100-1000 times lower than the observed fluxes, the light curves were normalized by dividing with the mean flux of the full light curve in order to make the comparison of flux distributions meaningful. The normalized flux distributions of the observed and the simulated light curves were then compared using the two-sided Anderson-Darling test and requiring at least a 95\% significance for the acceptance of a match. Figure~\ref{fig:distr} shows an example of one such comparison. In the upper panel, the normalized light curves are plotted on top of each other, the observed light curve in the >800 GeV band shown in green and the simulated light curve of the same energy range in pink. In the lower panel, the flux distributions are overplotted and the p-value of the Anderson-Darling test for this particular comparison is shown, in this case indicating a match. The number of matching distributions per  1000 sampled light curves was computed for each simulation scenario.

\begin{figure}[htbp]
\hspace*{-2cm}
\includegraphics[scale=1.4]{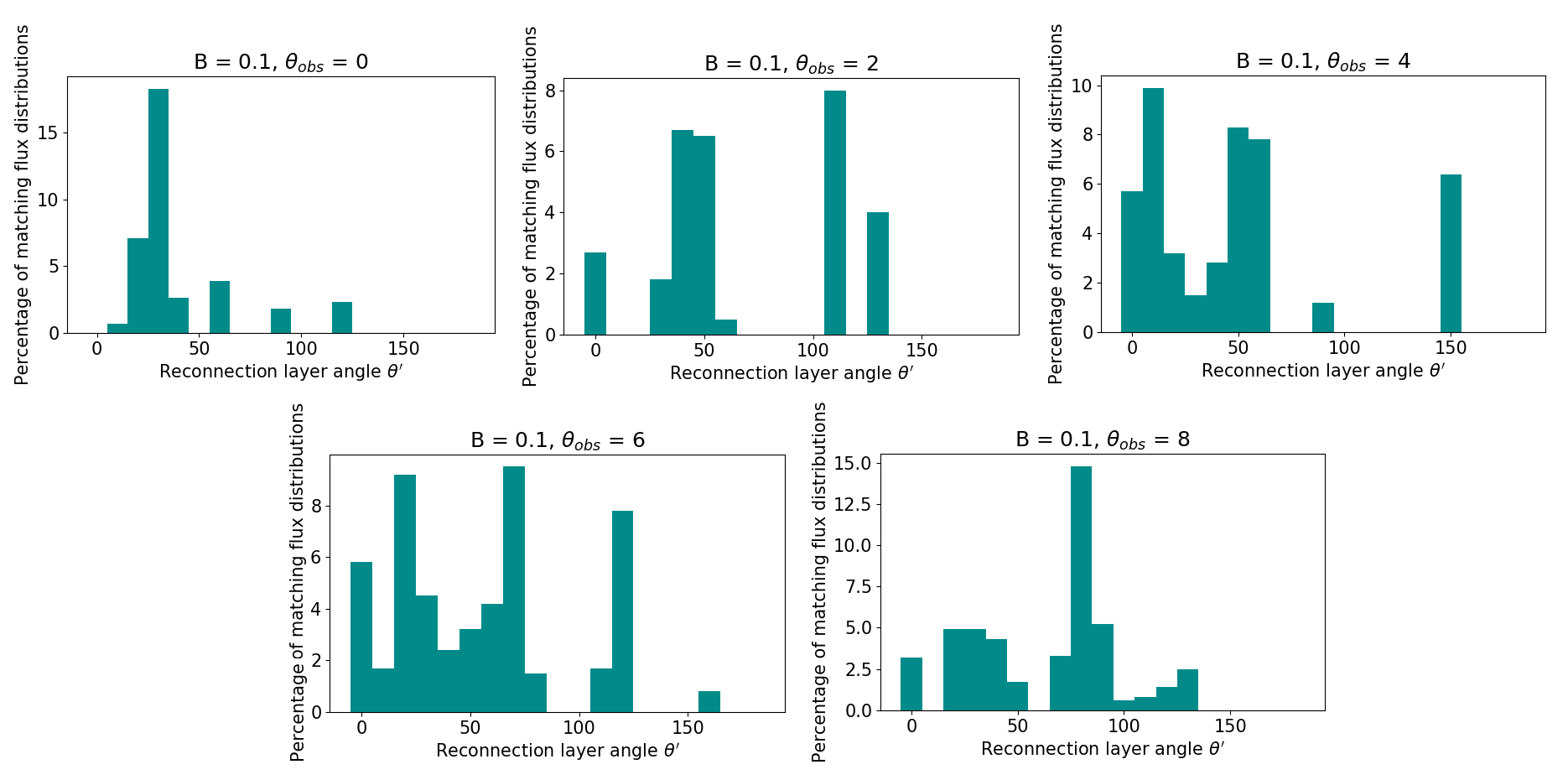}
  \caption{
  Histograms showing the reconnection layer angles for each jet orientation that produce matching distributions with the observed data. The number of matching simulations is given as a percentage of the 1000 different samplings (see text for details).}
  \label{fig:histo}
\end{figure}

We calculate the number of matches for each energy band, and because the number of matches in different energies were often not the same, the smallest number of matches out of the three energy bands was selected to be shown as a result for each simulated scenario. The percentages of matching distributions were plotted as histograms for each observation angle $\theta _{obs}$. Figure~\ref{fig:histo} shows examples of such histograms for each observation angle $\theta _{obs}$ = 0, 2, 4, 6, and 8 degrees. From these histograms we can see the range of reconnection layer angles that gives matching distributions depends on the observation angle.

\section{Discussion}
\label{discussion}

As can be seen from Figure~\ref{fig:histo}, already our preliminary results show that it is possible to find such combinations of jet parameters that produce simulated light curves that resemble the observations of Mrk 421, although the distributions are fairly wide. 
In \cite{Acciari2020}, the variability of Mrk 421 was also studied using a similar reconnection model. They only used one set of simulation parameters where $\sigma$ = 10  $\Gamma _{j}$ = 14 , and $\theta _{obs}$ = 2.1\degree. They analyzed both the fluxes and flux doubling time scales and found a range of reconnection layer angles of $30-90$\degree \ to be the best match with the observations. Comparing to our results of $\theta _{obs}$ = 2\degree \ we also find matches from layer angles of $30-60$\degree \ but also from 0\degree, 110\degree, and 130\degree. However, we note that we have not yet analyzed the variability time scales, which may make the range of matching parameters more narrow.


As it was pointed out in Section~\ref{distr}, the fluxes of our simulated light curves were often 100-1000 times lower than observed. Due to this mismatch, we decided to run a new set of simulations with tweaked input parameters to upscale the simulated fluxed but trying to maintain the variability and time scale behaviour of the simulations presented in this paper. The parameters that were changed were the bulk Lorentz factor and the reconnection layer half length L by increasing them by a factor of three. These will be presented in Jormanainen et al. (in prep.).

\section{Summary}
\label{summary}

In this paper, we present a first look of our study where we have compared simulated light curves obtained from relativistic magnetic reconnection models to observed data. We use one source, Mrk 421, to introduce our method. One part of our analysis method is the comparison of the flux distributions, which we present in this proceedings article. We find ranges of reconnection layer angles for each jet orientation that produce matching flux distributions with our observed data. In our future paper, we will describe our full analysis method that we aim to use also for other sources in different time scales and energies.

\clearpage
\bibliographystyle{ICRC.bst}
\bibliography{icrc.bib}

%
%
%

\end{document}